\documentclass[preprint,amsmath,nofootinbib]{revtex4}
\usepackage{graphics}

\begin{document}

\title{Predictions   and  Observations   in   Theories  with   Varying
  Couplings}             \author{C.              Armend\'ariz-Pic\'on}
\email{armen@oddjob.uchicago.edu} \affiliation{Enrico Fermi Institute,
  Department of Astronomy and Astrophysics, \\ University of Chicago.}

\begin{abstract}
  We  consider a toy  universe containing  conventional matter  and an
  additional real  scalar field, and  discuss how the  requirements of
  gauge  and  diffeomorphism   invariance  essentially  single  out  a
  particular set of theories which  might describe such a world at low
  energies.   In these  theories, fermion  masses and  $g$-factors, as
  well  as  the  electromagnetic  coupling  turn to  be  scalar  field
  dependent; fermion  charges and the gravitational  coupling might be
  assumed to  be constant.  We then  proceed to study the  impact of a
  time variation of the scalar field on measurements of atomic spectra
  at  high  redshifts.   Light   propagation  is  not  affected  by  a
  sufficiently slow change of the fine structure constant, but changes
  of  the  latter  as  well   as  variations  of  fermion  masses  and
  $g$-factors  do affect  the  observed atomic  spectra.  Finally,  we
  prove the independence of  these predictions on the chosen conformal
  frame, in  a further  attempt to address  differing views  about the
  subject expressed in the literature.
\end{abstract}

\maketitle

\section{Introduction and Summary}
Recent  analysis of  distant quasar  spectra  seem to  imply that  the
constants   of   nature  are   changing   with  time   \cite{Webbetal,
  Ivanchiketal}. Many different specific  models have been proposed to
explain such eventual changes, and most of them involve a scalar field
governing     the    value    of     the    constants     of    nature
\cite{JordanBransDicke,Magueijo,MoffatClayton}.      On     completely
different  measurements rests  the  solid evidence  that the  universe
contains  a component presently  driving accelerated  cosmic expansion
\cite{accelerating}.  Yet again, time-evolving scalar fields have been
widely considered as an alternative to explain the origin of late time
cosmic  acceleration  \cite{Carroll}.   Scalar  fields play  indeed  a
prominent  role in  modern  cosmology.  Their  importance arises  from
their  simplicity, their  power  to explain  many seemingly  unrelated
different  problems and  their  ubiquitous appearance  in theories  of
fundamental  particle  interactions.   However,  although  substantial
progress has  been achieved in  the past, it  is fair to say  that the
latter are far from being able to make concrete (realistic) low-energy
predictions.   Given  the possibility  that  the  universe contains  a
scalar field in addition to conventional matter, and given the to some
extent lack of theoretical guidance, it is important to ascertain what
is a generic  consequence of the existence of such  a scalar field and
what is not.

In Section \ref{sec:action} of this  paper, we assume that in addition
to ordinary matter there exists a single real scalar field potentially
relevant  at   low  energies,  and  construct   the  ``most  general''
low-energy theory that one might expect from basic symmetry principles
like  diffeomorphism and  gauge invariance.   The  resulting effective
action  has the  form one  would  expect from  general relativity  and
electromagnetism with  the exception that all  ``constants'' of nature
(gravitational  and electromagnetic  couplings,  magnetic moments  and
fermion masses as  well) can depend on the value of  the scalar field. 
Theories  of this  form  have  been long  advocated  as a  theoretical
framework for phenomenological  studies of gravity \cite{Damour}.  Our
construction is useful not only  because it delimits up to what extent
they are general, but also because  it shows the precise nature of the
assumptions made  during the derivation  of the effective  action, and
what could be different if any of the assumptions were relaxed.

The  effective  action  found   in  Section  \ref{sec:action}  can  be
considerably  shortened.    In  Section  \ref{sec:simplifications}  we
simplify  its   form  by   ``renaming''  fields  and   couplings.   In
particular, some of  of the above mentioned varying  quantities can be
assumed to be  constant without loss of generality.   We argue that in
the  present   context  the  most   natural  choice  is  to   fix  the
gravitational constant.   In order  to do so,  one needs to  perform a
``conformal transformation'', which is  nothing else than a particular
field  redefinition that  involves the  metric field.   Whereas  it is
commonly  accepted   that  field  redefinitions  do   not  affect  the
``meaningful'' predictions of any  theory, i.e.  the predictions which
can  be verified  in an  experiment, the  special role  played  by the
metric  in general  relativity  has  caused a  long  debate about  the
physical  equivalence  of  two  actions  that differ  by  a  conformal
transformation \cite{FaGuNa}.  Although our  purpose is not to provide
a general proof of such  equivalence, we have considered worthwhile to
show in Appendix \ref{sec:appendix}  with an explicit example that the
predictions  made by two  conformally related  actions are  indeed the
same.

Once the effective action has been cast in its simplest form, the next
goal of the paper is to  state some of the predictions that the theory
makes.  A priori, the presence of the additional scalar field could be
responsible  for  strong   violations  of  the  equivalence  principle
\cite{Carroll2},   and   therefore,   the  different   field-dependent
quantities in our theory  have to obey strong experimental constraints
\cite{Will,Damour}. We shall simply  assume that changes in the scalar
field might induce  variations in the couplings and  parameters of the
effective action, without making further suppositions about the nature
of these changes.  In order to subsequently study the phenomenological
implications  of these  changes,  it  is then  important  to cast  any
prediction  in  an ``experimentally''  meaningful  way.   Most of  the
experimental evidence about time  variation of the constants of nature
\cite{Webbetal,     Ivanchiketal}      and     cosmic     acceleration
\cite{accelerating} stems from  measurements of photon spectra emitted
by distant objects.   In fact, if the constants  of nature are allowed
to vary in time, it is natural to expect a larger departure from their
present  values the further  one looks  back in  the past.   Hence, in
Section \ref{sec:spectra} we  focus on light emission by  atoms and on
the propagation  of photons  in an expanding  universe.  It  turns out
that if the change of  the electromagnetic coupling is ``slow'', light
propagation is  not affected  by changes in  any of the  other varying
parameters.   However, because  in  our theory  frequencies in  atomic
transitions depend on fermion masses, the electromagnetic coupling and
the magnetic  moment of  the electron, a  careful analysis  shows that
ratios  of frequencies  of atomic  spectral lines  also depend  on all
those  parameters.   Therefore, measurements  of  deviations from  the
expected  values could  a priori  be due  not only  to changes  in the
electromagnetic  coupling,  but also  to  variations  of  some of  the
remaining ``constants'' of nature.   This fact might have implications
in the  analysis of experimental  data suggesting a  time-varying fine
structure constant \cite{Webbetal}.

\section{The low energy effective action}
\label{sec:action}

According  to the  present paradigm  in physics,  at  sufficiently low
energies nature can be described by an effective action which accounts
for  its low energy  degrees of  freedom, and  is invariant  under its
symmetries.  The low energy  effective action consists of an expansion
in the derivatives of the fields (particles) observed at low energies.
The higher  the amount of field  derivatives, the more  the effects of
the  corresponding  terms  are   expected  to  be  suppressed  in  the
predictions of the theory.

As  mentioned in  the  introduction, recent  experiments suggest  that
there  is an  additional  component  in the  universe  which seems  to
mediate  non-electromagnetic interactions.   The question  we  want to
answer is: Assuming  that this component is a  real scalar field, what
is the most  general effective action we might  expect to describe our
world? For our purposes, at the energies we are considering, the world
consists  of electrons,  neutrons  and protons  which  are subject  to
electromagnetic and gravitational interactions.  They are respectively
described by spin-1/2 fields $\psi_f$  (where $f$ runs through $e$ for
electrons, $p$ for protons and $n$ for neutrons), by a massless spin-1
field   $A_\mu$   (the   photon)   and  a   massless,   spin-2   field
$g_{\mu\nu}=g_{\nu\mu}$ (the graviton).   The extra ingredient we want
to  consider is  a scalar  field $\phi$  which might  also  mediate an
additional form of ``gravitational'' interaction.

The  low  energy  effective  action  has to  be  invariant  under  the
symmetries  one wants to  impose on  the system.   In our  case, these
symmetries are intimately related to  the field content of the theory. 
In fact,  it seems  that the only  way of consistently  describing the
electromagnetic  field $A_\mu$  is by  incorporating  gauge invariance
into  the  theory,  and  analogously,  the only  way  of  consistently
describing a massless spin-2  field is by incorporating diffeomorphism
invariance   \cite{Weinberg}.  By   definition,  under   $U(1)$  gauge
transformations the fields transform according to
\begin{subequations}\label{eq:gauge-trf}
\begin{eqnarray}
  \phi &\to& \phi \\
  \psi_f &\to& \exp\left(i q_f\epsilon\right) \psi_f\\
  A_\mu &\to& A_\mu+\partial_\mu \epsilon \\
  g_{\mu\nu}&\to& g_{\mu\nu},
\end{eqnarray}
\end{subequations}
where  $\epsilon$ is  the gauge  parameter and  the charges  $q_f$ are
constants.    Note  that   a  \emph{real}   scalar  field   cannot  be
electrically     charged.      Under     diffeomorphisms     $x^\mu\to
\tilde{x}^\mu=\tilde{x}^\mu(x^\nu)$ the fields transform according to
\begin{subequations}
\label{eq:coordinate-trf}
\begin{eqnarray}
  \phi &\to& \phi \\
  \psi_f &\to& \psi_f \\
  A_\mu &\to& \frac{\partial x^\alpha}{\partial \widetilde{x}^\mu}A_\alpha \\
  g_{\mu\nu} &\to& \frac{\partial x^\alpha}{\partial \widetilde{x}^\mu}
  \frac{\partial x^\beta}{\partial \widetilde{x}^\nu}g_{\alpha\beta}.
\end{eqnarray}
\end{subequations}
Fermions are  diffeomorphism scalars. Indeed, a  different approach is
needed  in order to  couple fermions  to gravity  \cite{spinors}.  For
that  purpose,  one  introduces  a  set of  four  orthonormal  vectors
$e^\mu{}_a$  at each  point  of the  spacetime manifold,  $e^{\mu}{}_a
e^\nu{}_b\, g_{\mu\nu}=\eta_{ab}$.  The vierbein $e^{\mu}{}_a$ is only
determined  by the  metric up  to local  Lorentz  transformations.  In
order to avoid the appearance of new degrees of the freedom associated
with  the vierbein,  Lorentz transformations  are postulated  to  be a
local symmetry  of the theory.  Under  (local) Lorentz transformations
$\Lambda^a{}_b(x)$, the different fields transform according to
\begin{subequations}
\label{eq:lorentz-trf}
\begin{eqnarray}
\phi &\to& \phi \\
  \psi_f &\to& U(\Lambda)\psi_f \\
  A_\mu &\to& A_\mu \\
  e^\mu{}_a & \to& \Lambda_a{}^b e^\mu{}_b \\  
  g_{\mu\nu}&\to& g_{\mu\nu},
\end{eqnarray}
\end{subequations}
where $U$  is the  Dirac spinor representation  of the Lorentz  group. 
Note   that   we   postulate   invariance  under   arbitrary   Lorentz
transformations,    including    ``time   reversal''    $\mathcal{T}$:
$e^\mu{}_0\to  -e^\mu{}_0$   and  ``space  inversion''  $\mathcal{P}$:
$e^\mu{}_i\to -e^\mu{}_i$.

The  transformation  properties of  the  different fields  essentially
determine  the precise  nature  of the  particles  we are  describing,
allowing  us  to  distinguish  between ``fermions'',  ``photons''  and
``gravitons''.   The  next  step  is  to construct  a  (local)  action
functional of the fields  $\phi$, $A_\mu$ $g_{\mu\nu}$ (and $e^\mu_a$)
which  is invariant  under  the transformations  (\ref{eq:gauge-trf}),
(\ref{eq:coordinate-trf}) and (\ref{eq:lorentz-trf}). At low energies,
for  slowly varying  fields, terms  in the  Lagrangian with  the least
possible number  of derivatives give the  largest contributions. Thus,
we shall organize  the Lagrangian as an expansion  in the total number
of derivatives.  The least possible number of derivatives is two for a
dynamical  boson  $\sim(\partial \phi)^2$,  and  one  for a  dynamical
fermion  $\sim\overline{\psi}\partial\psi$.  We  want  to ascribe  the
same  weight  to the  two  kinetic terms,  and  this  can be  formally
accomplished by  assigning half an ``effective  derivative" to fermion
fields,  as  in  supersymmetric  effective theories  \cite{GSW}.   Our
theory  is  certainly  not  supersymmetric,  but  since  the  previous
argument  just relies  on  dimensional reasons  this procedure  should
still be a consistent way of organizing the long-wavelength expansion.
Hence, we  shall include in  our effective Lagrangian terms  with only
two  ``effective derivatives", i.e.   terms where  the number  of real
derivatives  plus one half  the number  of fermion  fields is  less or
equal two.   Furthermore, for the purposes of  studying atomic spectra
it will suffice to consider fermion bilinears.

\paragraph{Zero derivatives}
Gauge, Lorentz and diffeomorphism symmetries only allow the terms
\begin{subequations}
\label{eq:zero-derivatives}
\begin{eqnarray}
&c(\phi) \\ 
&\overline{\psi}_f m_f(\phi)\psi_f, 
\end{eqnarray}
\end{subequations}
where  $c$ and  $M_f$  are  arbitrary functions  of  the scalar  field
$\phi$.  A term $A_\mu A^\mu$  is not gauge invariant, and Fermi terms
$(\overline{\psi}\psi)(  \overline{\psi}\psi)$  are  left out  because
they  are  not fermion  bilinears.   The  Dirac  adjoint is  given  by
$\overline{\psi}=\psi^{+}\gamma^0$, where $\gamma^a$ is a set of Dirac
matrices, $\{\gamma^a,\gamma^b\}=2\eta^{ab}$. Observe  that the only 5
possible independent  fermion bilinears are  $\overline{\psi} M \psi$,
where   $M$   is   proportional   to  $1$,   $\gamma^5$,   $\gamma^a$,
$\gamma^a\gamma^5$  and   $[\gamma^a,\gamma^b]$.   Because  electrons,
protons and  neurons have  different charges, gauge  invariance forces
fermion bilinears to contain only one type of fermion.

\paragraph{One derivative}
Possibly the  only way to  insure the invariance  of our theory  is to
consider covariant  derivatives.  Up to multiplications  with terms in
eqs.  (\ref{eq:zero-derivatives}), the allowed  invariant combinations
are
\begin{subequations}
\label{eq:one-derivative}
\begin{eqnarray}
  i\,\overline{\psi}_f D\hspace{-.66em}/_f\,  \psi_f \\
  i\,\overline{\psi}_f \Gamma^\mu\psi_f \partial_\mu \phi
  \label{eq:scalar-potential}\\
  i\,\overline{\psi}_f [\Gamma^\mu,\Gamma^\nu] \psi_f F_{\mu\nu},
\end{eqnarray}
\end{subequations}
where $F_{\mu\nu}\equiv \partial_\mu  A_\nu-\partial_\nu A_\mu$ is the
Maxwell  tensor  and  $\Gamma^\mu=e^\mu{}_a \gamma^a$.  The  covariant
derivative of a fermion is
\begin{equation}\label{eq:fermion-covariant}
  D\hspace{-.66em}/_f=\Gamma^\mu\left(\partial_\mu+
    \frac{1}{2}\omega_\mu{}_{ab}\Sigma^{ab}-iq_fA_\mu\right),
\end{equation}
where  $\omega_\mu{}_{ab}$  is   the  (minimal)  spin  connection  and
$\Sigma^{ab}=\frac{1}{4}[\gamma^a,\gamma^b]$ are the generators of the
Dirac  representation of  the Lorentz  group  \cite{spinors}.  Without
loss of generality we can  assume that the covariant derivative of the
metric vanishes.   Otherwise, it had  to be expressible as  a properly
transforming  combination of the  fields in  the action  [for instance
$\nabla_\mu     \,g_{\nu\rho}=g_{\nu\rho}\partial_\mu\phi+\cdots$].    
Invariant terms  involving $\nabla_\mu \,g_{\nu\rho}$  [for instance $
g^{\nu\rho}\partial^\mu  \phi\, \nabla_\mu\, g_{\nu\rho}$]  could then
be cast as couplings that do not involve derivatives of the metric [in
our   example  $4\,\partial_\mu\phi   \partial^\mu\phi+\cdots$].   For
analogous reasons we  also set the torsion of  derivative operators to
zero.  A  non vanishing torsion would  manifest itself in  the form of
specific  couplings between  the different  fields, and  because these
have to be invariant under the considered symmetries, we will consider
them anyway.  One can also define the dual of the Maxwell tensor,
\begin{displaymath}
{}^*\!F_{\mu\nu}\equiv 
\det (e_\alpha{}^a)\,
\epsilon_{\mu\nu\rho\sigma}
F^{\rho\sigma}, 
\end{displaymath}
which  is  a tensor  under  diffeomorphisms,  but  changes sign  under
$\mathcal{P}$   and    $\mathcal{T}$.    A   term    proportional   to
$\overline{\psi}[\Gamma^\mu,\Gamma^\nu]\psi\,    ^*\!F_{\mu\nu}$    is
excluded because  it violates  the latter symmetries.   A contribution
proportional  to $\nabla_\mu  (\overline{\psi}_f  \Gamma^\mu \psi_f)$,
which vanishes on-shell if fermion  flavor is conserved, can be traded
for  a  term  proportional  to  (\ref{eq:scalar-potential})  after  an
integration by parts.

\paragraph{Two  derivatives}
Up to  a multiplication by a  scalar function the  only allowed scalar
combinations of two field derivatives are
\begin{subequations}
\label{eq:two-derivatives}
\begin{eqnarray}
  R\\
  F_{\mu\nu}F^{\mu\nu}\\
\partial_\mu\phi \partial^\mu \phi, \label{eq:phi-squared}
\end{eqnarray}
\end{subequations}
where      $R$     is      the      scalar     curvature.       Again,
$F_{\mu\nu}{}^*\!F^{\mu\nu}$   is   excluded   by   $\mathcal{T}$   or
$\mathcal{P}$ invariance.  The expression $\nabla^\mu\partial_\mu\phi$
can be  turned into  (\ref{eq:phi-squared}) [up to  a $\phi$-dependent
coefficient]  by   an  integration  by  parts.   Terms   of  the  form
$\overline{\psi}_f\Gamma^\mu\psi_f  F_\mu{}^\nu\partial_\nu  \phi$  or
$\overline{\psi}_f\Gamma^\mu\gamma^5\psi_f\,
{}^*\!F_\mu{}^\nu\partial_\nu  \phi$ are  expected  from a  connection
with  torsion;  they  are  not  included because  they  contain  three
effective derivatives.

We can  construct gauge and  coordinate invariant terms in  our action
functional by combining the previous building blocks into factors with
at most  two effective derivatives.   In addition, in order  to define
our action as a local field functional, we need a coordinate and gauge
invariant ``volume element'' to  integrate those invariant terms.  The
most general expression of this type is
\begin{displaymath}
  \sqrt{|\det [v_g(\phi,\partial\phi^2) g_{\mu\nu}+v_F(\phi)F_{\mu\nu}
    +v_{1}(\phi)\partial_\mu\phi \partial_\nu \phi
    +v_{2}(\phi)\nabla_\mu\partial_\nu\phi]|}\,d^4x, 
\end{displaymath}
which   yields   a    generalization   of   the   Born-Infeld   action
\cite{BornInfeld}.   By  expanding  the  square  root  in  powers  one
recovers to lowest order terms proportional to $F_{\mu\nu}F^{\mu\nu}$,
$(\partial  \phi)^2$ and $\nabla^\mu\partial_\mu\phi$  plus additional
higher  order derivatives.   Because  we have  already considered  the
lowest    order    ones   as    part    of    our   building    blocks
(\ref{eq:two-derivatives}), and  we keep  only up to  two derivatives,
without loss of  generality we can set $v_F=v_1=v_2=0$.   For the same
reasons, we can  assume that $v_g$ only depends  on $\phi$.  The final
(unsimplified) action thus reads
\begin{eqnarray}\label{eq:most-general}
  S=\int d^4x \sqrt{-g}\, v_g^2(\phi)\Bigg\{-\frac{B_g(\phi)}{16\pi}R
  -\frac{B_F(\phi)}{16\pi}F_{\mu\nu}F^{\mu\nu}
  +\frac{B_\phi(\phi)}{2}\partial_\mu\phi\partial^\mu \phi 
  -V(\phi)+ \nonumber \\
  +B_f(\phi)\overline{\psi}_f\left[iD\hspace{-.66em}/_f-m_f(\phi)\right]\psi_f
  -i \overline{\psi}_f\Gamma^\mu \psi_f \partial_\mu r_f(\phi)
  +i\frac{q_f h_f(\phi)}{16 m_f(\phi)}
  \overline{\psi}_f[\Gamma^\mu,\Gamma^\nu]\psi_f F_{\mu\nu}
  \Bigg\}.
\end{eqnarray}
Note that all the terms containing boson differentiations only involve
partial  derivatives,  and  hence,   are  independent  of  the  way  a
connection is defined.
 
In order to delimit the validity of our derivation, let us summarize
all the assumptions made:
\begin{itemize}
\item Locality
\item Diffeomorphism invariance (in 4 spacetime dimensions)
\item  Invariance  under  $\mathcal{P}$  or  $\mathcal{T}$  and  local
  Lorentz transformations
\item $U(1)$ gauge invariance
\item  Lowest order  in derivative  expansion (at  most  two effective
  derivatives)
\item Bilinear in fermion fields
\end{itemize}
Departures  from   four  dimensional  coordinate   invariance,  as  in
brane-world      models      \cite{brane-worlds}     (see      however
\cite{KhouryZhang}),  and  certain  theories  where  higher-derivative
terms become large \cite{k} are  obvious examples that do not fit into
our  framework.   Our  derivation  is  rather  to  be  regarded  as  a
conservative  attempt to  delimit  a basic,  but  still quite  general
reasonable  set  of  theories  to  focus  on.   In  fact,  the  action
(\ref{eq:most-general}) is general enough to accommodate scalar-tensor
theories  of gravity  \cite{JordanBransDicke}, Bekenstein's  theory of
varying $\alpha$ and its revivals \cite{Dicke2,Magueijo}, and even the
long-wavelength  limit  of  bimetric  theories  \cite{MoffatClayton}.  
Finally,   let   us  point   out   that   the   arguably  only   known
``ultraviolet-complete'' theory of  quantum gravity, string theory, is
expected to have an action  of the form (\ref{eq:most-general}) as its
low-energy limit \cite{Damour}.

\section{Simplifications}
\label{sec:simplifications}

The  action (\ref{eq:most-general})  can be  considerably  simplified. 
First, the  overall factor  $v_g$ can be  absorbed into  the remaining
$\phi$  dependent functions  $B_g$, $B_F$,  etc.  Next,  by performing
field redefinitions,  some of the $B$  functions can be  assumed to be
constant.    Specifically,  by  introducing   the  new   scalar  field
$d\widetilde{\phi}=\sqrt{B(\phi)}\,d\phi$   one  can   always  choose
$\widetilde{B_\phi}=\pm        1$,        and       by        defining
$\widetilde{\psi}_f=\sqrt{B_f(\phi)}\psi_f$   one   can  also   choose
$\widetilde{B_f}=\pm1$. In addition, by the field redefinition
\begin{equation}\label{eq:conformal-transformation}
\widetilde{g}_{\mu\nu}=\Omega^2 g_{\mu\nu},
\end{equation}
where  $\Omega^2=B_g(\phi)$, one  can  set $\widetilde{B_g}(\phi)=1$.  
Under    the    transformation    (\ref{eq:conformal-transformation}),
sometimes   called   a   ``conformal   transformation'',   canonically
normalized ($B_f=1$) fermion masses transform according to
\begin{equation}\label{eq:conformal-m}
  \widetilde{m_f}= \Omega^{-1} m_f.
\end{equation}
Thus,  instead of requiring  $B_g=1$ we  could set  one of  the masses
$m_f$  to  one  \cite{Dick}.    In  such  a  ``conformal  frame''  the
corresponding  fermion would  be minimally  coupled to  the ``metric''
$g_{\mu\nu}$.   However, since $B_F$  as well  as $h_f$  are invariant
under the redefinition (\ref{eq:conformal-transformation}),
\begin{equation}\label{eq:conformal-F}
  \widetilde{B_F}=B_F,\quad \widetilde{h_f}=h_f,
\end{equation}
and because  in general  there is no  reason to expect  the parameters
$m_f$ to be proportional to  each other, matter would be still coupled
to  the scalar  field.   We  therefore conclude  that  in the  present
framework $B_g=1$  is the  most convenient choice.   Consequently, the
following action is completely equivalent to (\ref{eq:most-general}),
\begin{eqnarray}\label{eq:simplified-action}
  S=\int d^4x \sqrt{-g} \Bigg\{-\frac{R}{16\pi}
  -\frac{B_F(\phi)}{16\pi}F_{\mu\nu}F^{\mu\nu}
  \pm\frac{1}{2}\partial_\mu\phi\partial^\mu \phi 
  -V(\phi)\pm \nonumber \\
  {}\pm\overline{\psi}_f\left[iD\hspace{-.66em}/_f-m_f(\phi)\right]\psi_f
  -i \overline{\psi}_f\Gamma^\mu \psi_f \partial_\mu r_f(\phi)
  +i\frac{q_f h_f(\phi)}{16 m_f(\phi)}
  \overline{\psi}_f F_{\mu\nu} [\Gamma^\mu,\Gamma^\nu]\psi_f
  \Bigg\},
\end{eqnarray}
where  the  fermionic  covariant   derivative  is  given  by  equation
(\ref{eq:fermion-covariant}).  Note that  symmetry principles alone do
not restrict the signs of the kinetic terms.

It  is important  to make  a  distinction between  the parameters  and
fields  that  enter the  action  (\ref{eq:simplified-action}) and  the
quantities one  measures in real  experiments.  Although we  have been
talking about the  ``metric'', the ``gravitational coupling'', fermion
``masses'' and so on, one  should realize that the quantities measured
in experiments  might be  different, even if  they are given  the same
name.   For  instance,  in  standard general  relativity  $g_{\mu\nu}$
determines  proper distances  and  times, and  hence, $g_{\mu\nu}$  is
generally  denoted  as the  ``metric''.   Because  the  metric is  not
invariant under  conformal transformations, doubts  about the physical
equivalence  of conformally related  actions have  been raised  in the
literature \cite{FaGuNa}.  In our  description of gravity however, the
``metric'' is just an additional field essentially on the same footing
as  the  remaining  ones  \cite{Dicke},  and its  precise  meaning  is
determined  by  its  couplings   to  them.   In  particular,  distance
measurements might hinge on other parameters, like fermion masses.  We
have  considered  worthwhile  to  illustrate  the  issue  in  Appendix
\ref{sec:appendix}, where we show  the conformal frame independence of
the outcome of a redshift measurement.

When we  later study the motion  of electrons around a  nucleus, it is
going to  be convenient  to have a  point particle description  of the
fermion    instead    of    the    field   description    in    eq.    
(\ref{eq:simplified-action}).   What  matters  is  how  a  fermion  is
accelerated in  the presence of  the $\phi$, $A_\mu$  and $g_{\mu\nu}$
fields, so our goal is to compute this acceleration.  We shall neglect
the effects of the fermion spin, so we shall ignore terms proportional
to    $[\gamma^a,\gamma^b]$.    Then,    by    varying   the    action
(\ref{eq:simplified-action})  with respect to  $\overline{\psi}_f$ one
gets  the  Dirac  equation  $[i\Gamma^\mu  (\partial_\mu-i  q  A_\mu-i
\partial_\mu  r)-m]\psi=0$,  where we  have  dropped  the  $f$ label.  
Observe  that   the  function   $r$  plays  the   role  of   a  scalar
``electromagnetic  potential''.    The  WKB  ansatz   $\psi=\Psi  \exp
iS(x^\mu)$,  where $\Psi$  is a  constant spinor,  yields $[\Gamma^\mu
(\partial_\mu S- q A_\mu- \partial_\mu r)+m]\Psi=0$, which implies the
on-shell condition
\begin{equation}\label{eq:on-shell}
  g^{\mu\nu} (\partial_\mu  S- q  A_\mu -\partial_\mu r)
  (\partial_\nu S- q  A_\nu-\partial_\nu r)=m^2.
\end{equation}
We  thus  identify  the  mechanical  four  momentum  of  the  particle
$p^\mu$ to be 
\begin{equation}
  p^\mu\equiv m \cdot  u^\mu=\partial_\mu S-q A_\mu-\partial_\mu r,
\end{equation}
and differentiating the on-shell condition (\ref{eq:on-shell}) we find
then that the force exerted on the particle is given by
\begin{equation}\label{eq:point-particle}
 u^\mu \nabla_\mu(m u_\nu)=q F_{\nu\mu} u^\mu+\partial_\nu m. 
\end{equation}
The first  term on the right hand  side is the Lorentz  force, and the
second is the ``fifth force'' described by Dicke \cite{Dicke}. Because
derivatives acting on scalars commute, the particle does not couple to
$r$. The resulting  equation of motion can be  derived from the action
principle
\begin{equation}\label{eq:fermion-action}
  S_f=-\int d\lambda\, \left\{m_f\sqrt{g_{\mu\nu}
      \frac{dx^\mu}{d\lambda}\frac{dx^\nu}{d\lambda}}
    +q_f A_\mu \frac{dx^\mu}{d\lambda}\right\},
\end{equation}
where $m_f$ and $A_\mu$ are  evaluated at the position of the particle
$x^\mu(\lambda$).  We could have  also included a term $\partial_\mu r
(dx^\mu/d\lambda)$ in the action (\ref{eq:fermion-action}), which is a
total derivative and hence does  not alter the equations of motion, in
agreement with our previous remark concerning the same fact. Note that
the behavior of the fermion mass $m_f$ under conformal transformations
(\ref{eq:conformal-m})    can    be   also    derived    from   eq.    
(\ref{eq:fermion-action}).

\section{Atomic spectra}
\label{sec:spectra}

A  significant amount  of information  about our  universe  stems from
measurements of spectra emitted by distant objects. In this section we
consider an ideal measurement  whereby an electron in an hydrogen-like
atom changes its quantum state  and thereby emits a photon of definite
frequency.  The  photon freely propagates to  an hypothetical observer
who measures its frequency by comparing it to the frequency of photons
emitted  by a  reference  ``atomic  clock'' at  the  observer's site.  
Hence, in  order to predict the  result of the measurement  we need to
know  $a)$ how  the frequency  of the  emitted photon  depends  on the
parameters of  our theory,  $b)$ how the  photon freely  propagates in
space,  and  $c)$ how  the  observer  compares  the frequency  of  the
incoming photon to the one of a photon emitted by the reference atomic
clock. In the following we address these points in that order.

\setcounter{paragraph}{0}
\paragraph{Emission}

Consider a  hydrogen-like atom consisting  of a nucleus of  mass $m_N$
and charge $Z\cdot q_p$ surrounded  by a single electron.  The nucleus
itself can be described  by the action (\ref{eq:fermion-action}), with
$m_f=m_N$ and $q_f=Z q_p$.  Because  our toy universe does not contain
nuclear forces, such  an atom could not possibly  exist, but this fact
is not essential  for our purposes.  We shall  assume that there exist
coordinates where  the universe looks flat,  homogeneous and isotropic
\cite{flat},
\begin{equation}\label{eq:metric}
ds^2\equiv g_{\mu\nu}dx^\mu dx^\nu=a^2(\eta)(d\eta^2-d\vec{x}^2).
\end{equation}
Therefore, it also follows that $\phi$ can only depend on time $\eta$,
and the parameters in our action might hence also be $\eta$-dependent.
Following  the  standard procedure  to  quantize the  non-relativistic
limit of eq. (\ref{eq:fermion-action})  in the presence of an external
electromagnetic  field sourced  by the  nucleus  \cite{Gasiorowicz} we
find that  the Hamiltonian of the electron  has degenerate eigenvalues
given by
\begin{equation}\label{eq:zeroth-order}
 E^0_{n,l}=-Z^2\alpha^2 \frac{a\, \mu}{2 n^2},
\end{equation}
where  $\mu\equiv  m_e  m_N/(m_e+m_N$)  is  the reduced  mass  of  the
electron-nucleus system and
\begin{displaymath}
\alpha\equiv \frac{q_e^2}{B_F}
\end{displaymath}
is the  fine structure constant. It  is important to  notice that this
result is valid  so long as the scale factor $a$  and the field $\phi$
can  be  regarded  as  constant  (we shall  later  discuss  when  this
assumption  is applicable).   By  expanding  the square  root  in eq.  
(\ref{eq:fermion-action}) to a higher order in the fermion and nucleus
velocities, one  gets a relativistic  correction \cite{Gasiorowicz} to
the lowest order result (\ref{eq:zeroth-order}),
\begin{displaymath}
  \Delta E^{r}_{n,l}=-Z^4\alpha^4 \frac{a\, \mu}{2}
\left(\frac{\mu^3}{m_e^3}+\frac{\mu^3}{m_N^3}\right)
\left(\frac{1}{n^3(l+1/2)}-\frac{3}{4n^4}\right).
\end{displaymath}
A more accurate treatment of the motion of the electron with the Dirac
equation  additionally  introduces  the  spin-orbit  coupling  of  the
electron magnetic moment to the magnetic field \cite{Gasiorowicz},
\begin{displaymath}
  \Delta E^{so}_{n,l}=Z^4\alpha^4 \frac{a \mu}{2} \frac{\mu}{m_e}
\frac{1+h_e}{2}
\frac{\scriptstyle\left\{\begin{array}{c} l \\ -l-1 \end{array}\right\}}
{n^3 l(l+1/2)(l+1)},
\end{displaymath}
where the upper and lower values in the bracket apply for an atom with
total angular momentum  $j=l\pm\frac{1}{2}$ respectively.  In order to
compute the frequency  of an emitted photon in  an electron transition
from a state  $(n_i,j_i,l_i)$ to a state $(n_f,j_f,l_f)$  one needs to
quantize the electromagnetic field  too.  The photon field (in Coulomb
gauge) $A_\mu$ is thereby expanded into modes
\begin{displaymath}
  \vec{A}=\sum_{k} \vec{\mathcal{A}}_{k} e^{-i k_\mu x^\mu} a_k+h.c
\end{displaymath}
where the  sum runs over  null vectors, $k_\mu k^\mu=0$.   A necessary
condition for  the emission  of a photon  of four-momentum  $k_\mu$ is
``energy conservation'', $E_i=E_f+k_0$.   Therefore, the possible time
components of the photon four-momentum are
\begin{equation}\label{eq:emission-frequencies}
k_0\approx  Z^2\alpha^2 \frac{a \mu}{2}\left\{A-Z^2\alpha^2
\left[2B+(C/2-3B) \frac{\mu}{m_e}+\frac{C}{2} \frac{\mu}{m_e} h_e
\right]\right\},
\end{equation} 
where
\begin{subequations}
\label{eq:coefficients}
\begin{eqnarray}
A&=&\frac{1}{n_f^2}-\frac{1}{n_i^2} \\
B&=&\frac{1}{n_f^3(l_f+1/2)}-\frac{3}{4n_f^4}
- \frac{1}{n_i^3(l_i+1/2)}+\frac{3}{4n_i^4}\\
C&=&\frac{\scriptstyle\left\{\begin{array}{c}
l_f \\ -l_f-1 \end{array}\right\}}{n_f^3 l_f(l_f+1/2)(l_f+1)}
-\frac{\scriptstyle\left\{\begin{array}{c}
l_i \\ -l_i-1 \end{array}\right\}}{n_i^3 l_i(l_i+1/2)(l_i+1)}
\end{eqnarray}
\end{subequations}
only  depend  on  the  atomic  transition,  and  where  all  remaining
parameters are  to be evaluated at  the time of emission.   So long as
the relative change in those parameters during a time interval $1/k_0$
is negligible small, the assumption that they are constant should be a
good  approximation.  Note  that, besides  of $\alpha$  and  the scale
factor $a$,  both the mass ratio  $\mu/m_e$ and the  $g$-factor of the
electron $g_e=2+2h_e$ enter emission frequencies\footnote{Perturbative
  corrections  in  $\alpha$   also  contribute  to  the  ``anomalous''
  magnetic moment of the electron \cite{KinoshitaLindquist}; we absorb
  them into  $h_e$.  Similarly,  the mass of  the nucleus  contains an
  $\alpha$-dependent  piece due  to  the electromagnetic  interactions
  between nucleons.}.

\paragraph{Propagation}

We  assume that  once the  photon  is emitted  by an  atom, it  freely
propagate  in space  until it  reaches the  observer.  Whereas  it was
necessary to  quantize the electromagnetic  field in order  to compute
the possible emission  frequencies, a classical consideration suffices
to   determine   its  propagation.    The   Maxwell   term   in  eq.   
(\ref{eq:simplified-action})    is     invariant    under    conformal
transformations,  so that  the scale  factor in  the  conformally flat
metric (\ref{eq:metric})  does not enter the equations  of motion with
sources set to zero,
\begin{equation}\label{eq:maxwell}
\partial_\mu \left[B_F\, \eta^{\mu\rho}  \eta^{\nu\sigma} 
F_{\rho\sigma}\right]=0.
\end{equation}
Here $\eta_{\mu\nu}$  is the Minkowski metric and  $x^0=\eta$ is still
the  conformal  time  in  the metric  (\ref{eq:metric}).   Introducing
electric  and  magnetic   fields  $E_i=F_{0i}$,  $B_i={}^*F_{0i}$  and
defining  $\vec{D}\equiv   B_F  \vec{E}$,  $\quad   \vec{H}\equiv  B_F
\vec{B}$  eq.  (\ref{eq:maxwell})  translates  into  the  macroscopic
``inhomogeneous''  Maxwell equations $  \operatorname{div} \vec{D}=0$,
$\operatorname{rot} \vec{H}-\partial \vec{D}/\partial \eta=0$, whereas
by  definition $\operatorname{div}  \vec{B}=0,\quad \operatorname{rot}
\vec{E}+\partial \vec{B}/\partial  \eta=0$.  Thus, the  problem we are
studying is physically equivalent to the study of light propagation in
Minkowski space  permeated by a medium with  time varying permittivity
$\epsilon= B_F$ and permeability  $\mu= 1/B_F$.  The propagation speed
of the  electromagnetic perturbations, the  speed of light, is  in our
dimensionless units  given by  $v= (\epsilon \mu)^{-1/2}=1$,  which is
constant,  regardless of  how  all  the couplings  in  our theory  are
evolving.

Because  Maxwell's equations  are linear  in the  fields and  space is
homogeneous  and isotropic,  we can  decompose the  fields  field into
plane  waves $\propto  \exp(i  \vec{k}\cdot \vec{x})$.   Both sets  of
equations can then be combined into
\begin{equation}\label{eq:propagation}
  \left(B_F^{1/2}\vec{E}\right)''+
  \left[\vec{k}^2+\frac{1}{2}\frac{B_F''}{B_F}
    -\frac{3}{4}\left(\frac{B_F'}{B_F}\right)^2\right] 
  \left(B_F^{1/2} \vec{E}\right)=0,
\end{equation}
where a prime  means a derivative with respect  conformal time $\eta$. 
We assume that changes in $B_F$  are slow enough for a WKB solution to
be a good approximation,
\begin{equation}\label{eq:solution}
  \vec{E}=\frac{\vec{\mathcal{E}} }{\sqrt{2\Re(\omega_k})}
  \exp\left(i\int_{\eta_{em}}^\eta 
    \omega_k(\widetilde{\eta}) d\widetilde{\eta}\right).
\end{equation}
Here,  $\vec{\mathcal{E}}$  is   a  constant  transverse  polarization
vector, $\vec{k}\cdot\vec{\mathcal{E}}=0$, and $\Re(\omega_k)$ denotes
the real part of the ``frequency''
\begin{displaymath}
  \omega_k(\eta)=\sqrt{\vec{k}^2+\frac{1}{2}\frac{B_F''}{B_F}
  -\frac{3}{4}\left(\frac{B_F'}{B_F}\right)^2}+\frac{i}{2}
  \frac{B_F'}{B_F}.
\end{displaymath}
In order to uniquely  solve eq.  (\ref{eq:propagation}) proper initial
conditions  are  needed.   Because  the  solution  (\ref{eq:solution})
should describe light emitted by our  atom we require that at the time
of emission it behave as $\exp(i k_0 \eta)$, where $k_0$ is any of the
frequencies   in  eq.   (\ref{eq:emission-frequencies}).    Under  the
assumption of slowly varying $B_F$  this condition fixes the length of
the                wave                vector               $\vec{k}$,
$\omega_{em}\equiv\omega_k(\eta_{em})\approx   k_0(\eta_{em})$.   When
the photon reaches the observer at time $\eta_{arr}$, its frequency is
given by  $ \omega_{arr}\approx \omega_k(\eta_{arr})$.   Only if $B_F$
changes either during emission or observation is the frequency line of
the emitted light shifted and  broadened.  For a slowly changing $B_F$
the dominant effect is the broadening, which is of the order
\begin{displaymath}
  \frac{\delta\omega_{em}}{\omega_{em}}\approx
  \frac{1}{\omega_{em}}
  \left[\left(\frac{B_F'}{B_F}\right)_{arr}-
    \left(\frac{B_F'}{B_F}\right)_{em} \right].
\end{displaymath} 
Hence, if the relative change of  $B_F$ during the period of the field
oscillations both at the  emission and observations is negligible, the
effect of varying  $B_F$ on the photon propagation  is negligible too,
regardless of  the overall total change  of $B_F$ between  both times. 
We     shall     hence    ignore     this     effect    and     assume
$\omega_{arr}=\omega_{em}=k_0$.

\paragraph{Observation}
When the photon finally reaches  the observer at time $\eta_{arr}$, he
or  she can compare  its frequency  $\omega_{arr}=k_0(\eta_{em})$ with
the  one  of  a photon  emitted  by  a  reference atomic  clock.   For
simplicity we  take the  reference clock to  be identical to  the atom
that emitted  the photon  (same values of  $Z$ and $m_N$).   Then, the
atomic  clock frequencies  $\omega_{clock}$  are still  given  by eq.  
(\ref{eq:emission-frequencies}), but the  different factors have to be
evaluated        at       the        time        of       observation,
$\omega_{clock}=k_0(\eta_{arr})$.   The observer determines  the ratio
of the frequency of the incoming photon to the frequency of the photon
emitted by the clock to be
\begin{equation}\label{eq:redshift}
  \frac{\omega_{arr}}{\omega_{clock}}=
  \frac{(a\, \mu\, \alpha^2)_{em}}{(a\, \mu \,\alpha^2)_{arr}}
  \cdot \frac{\left\{A-Z^2 \alpha^2
      \left[2B+(C/2-3B)\,\mu/m_e
            +\frac{C}{2} (\mu/m_e) h_e\right]\right\}_{em}}
          {\left\{A-Z^2 \alpha^2
              \left[2B+(C/2-3B)\,\mu/m_e
            +\frac{C}{2} (\mu/m_e) h_e\right]\right\}_{arr}},
\end{equation}
where subscripts indicate the time where the corresponding expressions
should be  evaluated (emission or arrival), and  the coefficients $A$,
$B$,  $C$  are  given  by  eqs.  (\ref{eq:coefficients}).   Expression
(\ref{eq:redshift})   is  the   predicted  outcome   of   a  frequency
measurement.

In  general  relativity $\mu$  and  $\alpha$  are  constants, and  the
overall    coefficient    $a(\eta_{arr})/a(\eta_{em})$    in    eq.    
(\ref{eq:redshift})  is the  redshift, which  we have  derived without
explicitly assuming that proper times are determined by the metric. In
the present context, the redshift $z$ is rather given by
\begin{displaymath}
  1+z\equiv \frac{a(\eta_{arr})\mu(\eta_{arr})}
  {a(\eta_{em})\mu(\eta_{em})}.
\end{displaymath}
In the  absence of fine-structure corrections  ($B=C=0$) variations of
$\alpha$  cannot   be  distinguished  from   redshifts.   However,  by
considering several transitions (several  values of $A$, $B$ and $C$),
information about  the values of $a\cdot\mu$,  $\alpha$, $\mu/m_e$ and
$h_e$ at different times can  be in principle extracted from frequency
measurements.   At present  $\alpha_0\approx 1/137$,  for  an hydrogen
atom   $(\mu/m_e)_0\approx    1-5\cdot   10^{-4}$   ,    the   leading
electromagnetic  contribution to  the  $g$-factor of  the electron  is
$h^{em}_e\approx     \alpha/\pi+\mathcal{O}(\alpha^2)$,     and    the
non-electromagnetic  contribution is limited  by $h_e\lesssim10^{-10}$
\cite{KinoshitaLindquist}.  The important point is that the right hand
side of  eq.  (\ref{eq:redshift}) might differ  from $(1+z)^{-1}$ even
if $\alpha$ is constant.

\section{Conclusions}
\label{sec:conclusions}
We  have attempted  to construct  the most  general  low-energy action
consistent with  basic field  content and symmetry  requirements under
the assumption  of the existence of  a ``light'' scalar  field.  Up to
redefinitions  of fields  and couplings,  these  requirements uniquely
determine the form  of the effective action. In  this framework, it is
always possible  to choose the gravitational coupling  and the fermion
charges to  be constant.  However, fermion masses  and $g$-factors, as
well as  the electromagnetic coupling strength  are generically scalar
field dependent, and hence, possibly time-varying.

Because most of the theory  parameters and couplings depend on $\phi$,
the observed  frequency of  a photon emitted  in an  atomic transition
depends on the values of these parameters both at the time of emission
and  observation.   Concretely,  the   outcome  of  such  a  frequency
measurement might  be used to determine  not only changes  in the fine
structure constant, but also in  the $g$-factor of the electron and in
appropriate   mass  ratios.   Light   always  propagates   along  null
geodesics,  and its  frequency is  not  influenced by  changes in  the
electromagnetic  coupling  strength  as  long  as  these  changes  are
``slow''.

We have also  addressed the issue about the  dependence of our results
on  the  choice  of  conformal  frame.   In the  context  of  our  toy
experiment, frequency measurements are conformal frame independent, as
expected.

\begin{acknowledgments}
  It is a  pleasure to thank Rachel Bean, Sean  Carroll and his group,
  Ben Craps, Reiner Dick,  Norval Forston, Carlo Graziani, Jeff Harvey
  and  Slava Mukhanov  for  useful discussions.   This  work has  been
  supported by the U.S.  DoE grant DE-FG02-90ER40560.
\end{acknowledgments}

\appendix

\section{Conformal frame independence}\label{sec:appendix}
Our  derivation of  the  ``redshift''  measured by  an  observer, eq.  
(\ref{eq:redshift}),  has assumed  that the  action  is given  by eq.  
(\ref{eq:simplified-action}).   In particular  we have  worked  in the
``Einstein''  conformal  frame,  where  the function  multiplying  the
scalar curvature is a  constant.  Our original ``Jordan-frame'' action
(\ref{eq:most-general}) actually  contained a field-dependent function
$B_g$ multiplying $R$, but we were  able to remove it by the conformal
transformation (\ref{eq:conformal-transformation}).   A question which
has  been  repeatedly discussed  in  the  literature \cite{FaGuNa}  is
whether  actions  that  differ   by  a  conformal  transformation  are
equivalent.   If  they  were  not,  our simplification  might  not  be
justified.  Certainly, two such actions are mathematically equivalent,
in the  sense that solutions to  the equations of motion  in one frame
are  mapped by  the  conformal transformation  into  solutions of  the
equations  of  motion  in  the conformally  related  frame\footnote{We
  assume   that  no   singularities  appear   in   the  transformation
  \cite{AlvarezConde}.}.  However,  this does not  automatically imply
that they are physically  equivalent, in the sense of ``experimentally
meaningful'' predictions being identical.  Indeed, because there is no
general  framework  to  formulate  how  ``experimentally  meaningful''
predictions are to be extracted from a a particular set of fields, the
issue cannot be addressed in full generality.

Our goal  is to show that  in the context of  redshift measurements in
our toy  universe, conformally related actions  are indeed equivalent,
as  might   seem  to   be  obvious  if   one  regards   the  conformal
transformation  merely as  a field-redefinition.   For the  purpose of
illustration,  consider  an  ``expanding''  universe where  the  scale
factor $a$ in eq. (\ref{eq:metric}) grows with time and fermion masses
are   constant.     In   a    conformally   related   metric,    eq.   
(\ref{eq:conformal-transformation}), the  scale factor $\widetilde{a}$
is given by
\begin{equation}\label{eq:conformal-a}
\widetilde{a}=\Omega \, a,
\end{equation}
and fermion  masses vary according to  eq. (\ref{eq:conformal-m}).  By
an  appropriate  choice  of   the  arbitrary  function  $\Omega$,  the
conformally  related  universe  might be  ``contracting''  (decreasing
$\widetilde{a}$).    A  priori,   one  expects   both   expanding  and
contracting  universes to  be  quite different.   Let us  nevertheless
study how  an observer could  determine whether a universe  expands or
contracts.   Recall  that  the  first experimental  evidence  for  the
expansion of the universe  was E.  Hubble's measurements of redshifted
galaxy  spectra.  As a  matter of  fact, our  predicted redshift,  eq. 
(\ref{eq:redshift}) shows  that the observed frequency  of the photons
is proportional  to $1/a$.  The conformally related  action predicts a
frequency which is given simply by replacing the parameters that enter
eq.  (\ref{eq:redshift}) by their analogues in the conformally related
frame,
\begin{equation}
  \widetilde{\left(\frac{\omega_{arr}}{\omega_{clock}}\right)}=
  \frac{(\widetilde{a}\, \widetilde{\mu}\, 
    \widetilde{\alpha}^2)_{em}}{(\widetilde{a}\, \widetilde{\mu}
    \,\widetilde{\alpha}^2)_{arr}}
  \cdot \frac{\left\{A-Z^2 \widetilde{\alpha}^2
      \left[2B+(\frac{C}{2}-3B)\widetilde{\mu}/\widetilde{m}_e
            +\frac{C}{4} 
            (\widetilde{\mu}/\widetilde{m}_e)
            \widetilde{h}_e\right]\right\}_{em}}
          {\left\{A-Z^2 \widetilde{\alpha}^2
              \left[2B+(\frac{C}{2}-3B)\widetilde{\mu}/\widetilde{m}_e
            +\frac{C}{4}(\widetilde{\mu}/\widetilde{m}_e)
            \widetilde{h}_e\right]
        \right\}_{arr}}.
\end{equation}
Because  of  eq. (\ref{eq:conformal-m})  the  reduced  mass scales  as
$\tilde{\mu}=\Omega^{-1}\mu$,    and    therefore,    using    eqs.    
(\ref{eq:conformal-a}),  (\ref{eq:conformal-F})  and  bearing in  mind
that $q_e=\widetilde{q}_e$ we find
\begin{equation}
\widetilde{\left(\frac{\omega_{arr}}{\omega_{clock}}\right)}=
\left(\frac{\omega_{arr}}{\omega_{clock}}\right).
\end{equation}
Both observers measure the same  frequency ratios, the two actions are
physically equivalent  at this  level.  Nevertheless, the  behavior of
the scale factor in the two frames appears to be completely different.
In  fact,  as we  have  seen,  even if  atomic  spectra  appear to  be
redshifted, in  some conformal frames  the universe might  actually be
``contracting'' \cite{HoyleNarlikar}.

\end{document}